\documentclass[aps,prl,twocolumn,showpacs]{revtex4}
\usepackage{epsfig}
\newcommand{\be}{\begin{equation}} 
\newcommand{\ee}{\end{equation}}
\newcommand{\bea}{\begin{eqnarray}} 
\newcommand{\eea}{\end{eqnarray}}

\newcommand{\gton}{\mathrel{\lower.9ex \hbox{$\stackrel{\displaystyle 
>}{\sim}$}}} 
\newcommand{\lton}{\mathrel{\lower.9ex \hbox{$\stackrel{\displaystyle 
<}{\sim}$}}}

\newcommand{\vp}{{\vec p}}

\newcommand{\calN}{{\cal N}}

\begin{document}
\title{The ``minimal'' viscosity and elliptic flow at RHIC}
\author{Denes Molnar}
\affiliation{Physics Department, Purdue University, 
525 Northwestern Ave, West Lafayette, IN 47907, USA\\
RIKEN BNL Research Center, Brookhaven National Laboratory, Upton, 
NY 11973, USA}

\date{\today}
\begin{abstract}
We show from covariant transport theory 
that, for a massless ideal gas equation of state, 
even a small shear viscosity
to entropy density ratio
$\eta \approx s/(4\pi)$ 
generates significant $15-30\%$ dissipative corrections to elliptic
flow for 
conditions expected in mid-peripheral ($b = 8$ fm) $Au+Au$ collisions at 
$\sqrt{s_{NN}} \sim 200$~GeV at RHIC. 
\end{abstract}

\maketitle

{\em Introduction.}
The goal of the heavy-ion physics is to study the 
properties of nuclear matter at extreme energy densities and temperatures.
Spectacular features of the data 
from the Relativistic Heavy Ion Collider (RHIC)
for gold-gold reactions at center-of-mass energies
$\sqrt{s_{NN}} \sim 100 - 200$ GeV
lead to the suggestion that the hot and dense matter created in the collision
exhibits perfect fluid behavior.
(A perfect fluid has vanishing shear and bulk viscosities 
and heat conductivity.)
The cornerstone of this conclusion was 
the success of ideal (Euler) fluid dynamics
in explaining the large second Fourier moment
of the azimuthal momentum distribution for fixed $p_T$, 
the so-called ``elliptic flow'' 
$v_2(p_T) \equiv \langle \cos 2\phi \rangle_{p_T}$ 
\cite{flowreview,Kolb:2000sd,Huovinen:2001cy,Teaneyhydro,Hiranohydro}, 
observed in non-central collisions.

General considerations based on quantum mechanics, on the other hand,
indicate a nonzero {\em lower} bound for shear viscosity.
For example\cite{GyulassyDanielewicz85}, combining the
time-energy uncertainty principle with kinetic theory gives 
$\eta/s \gton 1/15$, where
$s$ is the entropy density.
This result is also supported by calculations for $\calN=4$ supersymmetric
Yang-Mills theory via the 
gauge-theory - gravity $AdS$-$CFT$ duality conjecture.
Pioneering results predicted
$\eta/s \ge 1/(4\pi)$ \cite{SonLimit},
while recent calculations for an extended class of theories find a somewhat
lower bound
$\eta/s \ge 16/25 \times 1/(4\pi)$~\cite{etaslimit_revised}.
It is an open question whether either of these limits applies to QCD.
But the possibility is intriguing because even a small
$\eta/s \sim {\cal O}(1)/(4\pi)$ has 
significant dynamical effects in heavy-ion collisions.

Ideal hydrodynamics assumes 
local equilibrium throughout the evolution.
For nonzero transport coefficients, on the other hand, 
the system departs from local equilibrium,
leading to dissipative corrections.
If the system stays sufficiently close to local equilibrium, dissipation 
can be investigated via causal dissipative hydrodynamics, for example, 
Israel-Stewart theory \cite{IS,Muronga,BRW}. 
Solution techniques for viscous hydrodynamic equations, in
the minimally required 2+1 dimensions necessary for elliptic flow studies,
have been recently developed
and applied\cite{romatschkeviscous,songviscous,teaneyviscous}.
However, Israel-Stewart theory comes from
a truncation procedure\cite{IS,DeGroot} with uncontrolled errors 
(it lacks a small expansion parameter),
and therefore its region of validity is not known.
Moreover, the solutions are causal only in a region of hydrodynamic parameters,
and their stability is not guaranteed\cite{unstableIS}.

Here we utilize instead the fully causal and stable 
covariant parton transport theory \cite{Bin:Et,ZPCv2,nonequil,v2,hytrv2}
for which covariant algorithms have been available in full 3+1D 
for quite some time\cite{ZPC,MPC,XG}. 
A forerunner of this analysis \cite{hytrv2} considered large,
{\em constant} elastic $2\to 2$ 
gluon-gluon cross sections $\sigma \sim 45$ mb, and found a 
significant $20-30$\% reduction of elliptic flow due to dissipation.
Such dynamics gives an increasing $\eta/s$ with time\cite{DerekQM2004}. 
In contrast, here we study dissipation
for a ``minimal'' $\eta/s \approx 1/(4\pi)$ that is constant in time.
Preliminary results have been
reported already at~\cite{prelim}, 
and are confirmed here to be accurate.

{\em Covariant transport theory near the hydrodynamic limit.}
We consider here, as in Refs. 
\cite{ZPC,Bin:Et,ZPCv2,nonequil,v2},
the simplest but nonlinear
form of Lorentz-covariant Boltzmann transport theory
in which the on-shell phase space density $f(x,\vp)$,
evolves with an elastic $2\to 2$ rate as
\bea
p_1^\mu \partial_\mu f_1 &=& S(x, \vp_1) 
+ \frac{1}{\pi} \int\limits_2\!\!\!\!
\int\limits_3\!\!\!\!
\int\limits_4\!\!
\left(f_3 f_4 - f_1 f_2\right)
W_{12\to 34} \nonumber\\
&&\qquad\qquad\qquad\quad\times \  \delta^4(p_1{+}p_2{-}p_3{-}p_4)
 \ .
\label{Boltzmann_eq}
\eea
where the integrals are shorthands
for $\int_i \equiv \int d^3 p_i / (2E_i)$.
For dilute systems, $f$ would be the phasespace distribution of 
quasi-particles, while the transition probability $W = s (s-4m^2) d\sigma/dt$
would be given by the scattering matrix 
element.
Our interest here, on the other hand, is to study the theory 
{\em near its hydrodynamic limit}.

It is well known (Boltzmann's \emph{$H$-theorem}) that
(\ref{Boltzmann_eq}) drives the system towards a fixed point,
global equilibrium.
In the \emph{hydrodynamic limit} ($W \to \infty$),
the transport solutions approach local equilibrium
$f(x,\vp) = g \exp[(\mu(x) - p_\nu u^\nu(x))/T(x)]/(2\pi)^3$.
A systematic expansion in small gradients around equilibrium 
via the Chapman-Enskog procedure \cite{DeGroot}
gives the viscous hydrodynamic equations by Navier and Stokes.
However, this approximate theory is severely acausal. 
A causal formulation
proposed by Mueller\cite{Mueller} and later generalized by 
Israel and Stuart (IS) 
\cite{IS} retains 
certain second-order derivative terms, resulting in relaxation equations.
IS theory can also be recovered from transport
via the 14-moment expansion of Grad \cite{IS,DeGroot}.
The transport
coefficients, and the microscopic relaxation times for the dissipative fluxes 
in IS, 
are all given by the differential cross section $d\sigma/dt$.

{\em 
The key observation here is that one can use transport theory to solve 
causal viscous
hydrodynamics provided one dials in the equation of state (EOS) 
and transport
coefficients of interest.} In this case, the ``particles'' and
the specific ``interaction'' in the transport 
have no physical significance - they are only mathematical tools to reproduce
the desired dynamical equations.
Here we consider an ultrarelativistic gluon gas with $e = 3p$, applicable
to the high-temperature plasma in the early stages at RHIC.
In this case\cite{DeGroot,IS}, $\eta \approx 4T/(5 \sigma_{tr})$ and the shear stress relaxation
time is $\tau_\pi = 6\lambda_{tr}/5$, where $\sigma_{tr}$ and 
$\lambda_{tr}\equiv 1/(n\sigma_{tr})$
are the transport cross section and transport mean free path, respectively.
Note, for an isotropic cross section $\sigma_{tr} = 2\sigma_{tot} / 3$.

{\em Elliptic flow and $\eta/s$.} For the above conditions, 
\be
\frac{\eta}{s} \approx \frac{\eta}{4n} \approx \frac{T\lambda_{tr}}{5} =
\frac{T}{5 n \sigma_{tr}}
\label{etas}
\ee 
Assuming the system stays close to
local equilibrium, during the initial longitudinal (Bjorken) expansion
stage of the heavy-ion collision the density and temperature evolve as 
$n \sim 1/\tau$, $T(\tau) \sim \tau^{-1/3}$ where
$\tau\equiv\sqrt{t^2-z^2}$ is the longitudinal (Bjorken) proper time.
For a constant cross section, $\eta/s \propto \tau^{2/3}$ then increases with 
time. 

\begin{figure}[htpb]
\epsfysize=5cm
\epsfbox{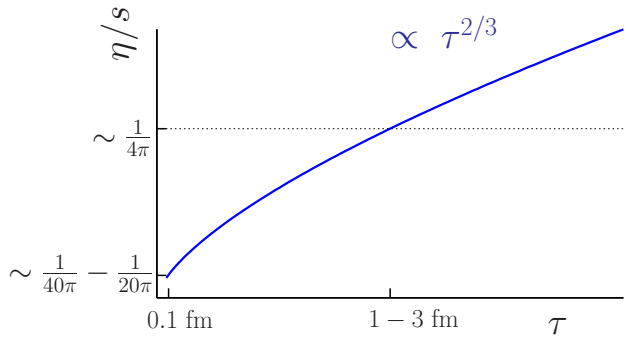}
\caption{Schematic evolution of $\eta/s$ for the calculation by 
Ref.~\cite{hytrv2} that considered a constant cross section. }
\label{fig:1}
\end{figure}

One might therefore think that Ref.~\cite{hytrv2} with constant 
$\sigma \sim 45$ mb overestimated dissipative effects in $Au+Au$ at RHIC 
energies. However,
even though $\eta/s$ grew in that calculation, its initial value was
really small (cf. Fig.~\ref{fig:1}). 
For the longitudinally boost invariant scenario 
assumed there,
$\lambda_{tr} = \tau / (\sigma_{tr}  dN/d\eta dx_T^2)$,
and with the parameters of that calculation\footnote{%
\label{IC}
$Au+Au$ at 
$b=8$ fm impact parameter, binary collision transverse profile,
$dN(b{=}8{\rm fm})/d\eta \approx 250$ 
(the corresponding maximum transverse density is 
$dN/d\eta dx_T^2 \approx 9.36$~fm$^{-2}$),
$T_0 = 0.7$ GeV, and Debye-screened cross section with
$\sigma_{tr} \approx 15$ mb.}
$\lambda_{tr}(\tau_0=0.1fm) \approx 7.1\times 10^{-3}$~fm(!),
i.e., $\eta/s \sim 1/(60\pi)$ at the very center of the collision.
The ratio is way below the conjectured bounds
even for the average density that is $2-3$ times smaller than the maximum.

We cross-check this important finding with the transport 
opacity\cite{v2}
\be
\chi\equiv \frac{\sigma_{tr}}{\sigma_{tot}} \langle n_{coll} \rangle
= \langle \int 
\frac{dz}{\lambda_{tr}({\bf x}_0+ z\hat{\bf n},\tau=\tau_0+z)}\rangle
\;\; \label{tropacity}
\ee
which is the number of collisions per particle weighted by
the transport cross section and averaged over 
initial coordinates and directions.
$\chi$ is dominated by the early and densest 
longitudinal expansion stage, during which $\lambda_{tr} \propto \tau$,
and thus
\be
\chi \approx \frac{\tau_0}{\langle\lambda_{tr}(\tau_0)\rangle} 
\int_0^{L} \frac{dz}{z+\tau_0}
\approx \frac{\tau_0}{\langle \lambda_{tr}(\tau_0)\rangle} 
\ln \frac{L}{\tau_0} \ .
\label{chi_const}
\ee
With $\chi \approx 21$ from the calculation and an estimate 
$L\sim 3-4$~fm for
the size of the reaction zone, {\em on average}
$\langle \lambda_{tr}(\tau_0=0.1fm)\rangle \approx 1.5\times 10^{-2}$ fm.
This is about 2.5 times larger than 
the value estimated for the collision center, as expected.

From (\ref{etas}) we then find that for the situation 
in Ref.~\cite{hytrv2}, $\eta/s$ evolves with time schematically
as shown in Fig.~\ref{fig:1}. During 
the first few fermis of the evolution relevant for the buildup of
elliptic flow~\cite{ZPCv2,hytrv2}, 
the system stays closer to equilibrium than would be allowed by a
``minimal'' viscosity because the transport
mean free path (or equivalently, the scattering rate) 
is not limited by any quantum bound.
The situation only changes after $\tau \sim 1-3$~fm, when the transport
mean free path becomes large enough to reach
$\eta/s \approx 1/(4\pi)$.

To study elliptic flow for constant $\eta/s$, 
we therefore start with an initially 
modest transport cross section 
and increase it with time as
$\sigma_{tr}(\tau) = \sigma_0 [\tau/(0.1 fm)]^{2/3}$.
Such growth is already encoded in perturbative QCD, provided we ignore
the running of the coupling:
$\sigma_{tr} \approx (18\pi\alpha_s^2/ s) \ln (s/\mu_D^2) 
\sim (\pi \alpha_s^2/T^2) \ln (18/g^2)$, 
where $\mu_D \sim gT$ is the Debye mass.
The initial conditions for $Au+Au$ at $\sqrt{s_{NN}} = 200$~GeV at impact
parameter $b = 8$~fm 
(see footnote \ref{IC} [31])
and the numerical solution technique MPC\cite{MPC} are the same as in 
Ref.~\cite{hytrv2}.
For numerical convenience we uniformly set $\sigma_0 \approx 2.7$~mb,
which ensures that on average $\eta/s \approx 1/(4\pi)$ in the system.
In the center of the collision zone, (\ref{etas}) gives a lower value
$\eta/s \approx 0.4/(4\pi)$, but that is compensated by the 
increase of
$\eta/s$ with decreasing density as we go outward. With the
average density $\langle n\rangle \sim n_{max} / 2.5$ estimated
earlier from (\ref{chi_const}), 
on average $4\pi\eta/s \approx 1$. A cross-check with the transport 
opacity $\chi \approx 16$ obtained from the growing-cross-section
calculation $\sigma_{tr}(\tau) \propto \tau^{2/3}$ gives
\be
\chi \approx \frac{3\tau_0}{2\langle\lambda_{tr}(\tau_0) \rangle} 
\left(\frac{L}{\tau_0}\right)^{2/3} \ ,
\label{chi_grow}
\ee
i.e., $\lambda_{tr}(\tau_0{=} 0.1fm) \sim 0.09-0.11$~fm and
$4\pi \eta/s \sim 0.8-1$. 

The above choice of $\sigma_0$ implies $\sim 5-10$ times higher 
two-body rates than perturbative QCD estimates. 
For our purposes this is not a problem,
the rates are simply adjusted to reproduce the desired $\eta/s$. 
With the inclusion of radiative $3\leftrightarrow 2$ processes\cite{XGetas},
even perturbative rates could generate a small $\eta/s \sim 0.1$.

With $2\to 2$ scattering, particle number is conserved
and thus $s = n (4 - \mu/T)$ where $\mu$ is the chemical potential.
Therefore, (\ref{etas}) acquires a small
relative correction
$\Delta s / s(\mu{=}0) 
= (1/4) \ln (n_{eq}/n) = (1/4) \ln [gT^3/(\pi^2 n)]$ 
logarithmic in density. 
For our gluon gas initial conditions ($g = 16$), it is about $\sim (-6)$\%
at the center of the collision, while $\sim 20$\% for a low 
$n = n_{max}/ 3$.
During the longitudinal expansion stage,
the correction stays roughly constant because
the dilution is largely compensated
by cooling, $n T^3 \approx const$.
Dissipation of course still generates entropy
because the temperature drops slightly slower\cite{GyulassyDanielewicz85} 
than $T \propto \tau^{-1/3}$,
but that effect cannot be very large for the system to stay near equilibrium.
With $4\pi\eta/s \sim 0.8 -1$ initially, we have a cushion
for entropy production during later evolution.
Therefore, we conclude that the averge $\eta/s$ is set to the desired 
$\eta/s = 1/(4\pi)$ within about $20$\% in the calculation.

\begin{figure}[htpb]
\epsfysize=6cm
\epsfbox{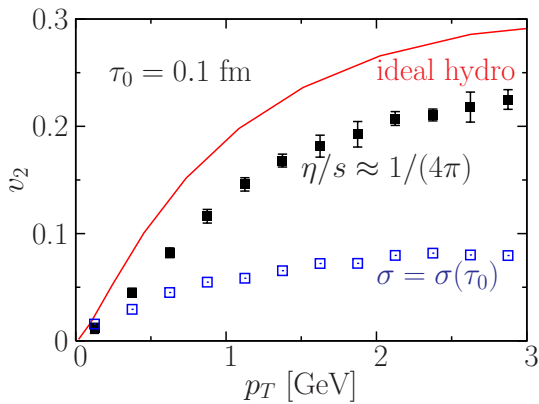}
\caption{Differential elliptic flow $v_2(p_T)$ as a function of $p_T$
in $Au+Au$ at $\sqrt{s_{NN}}=200$ GeV and $b=8$ fm at RHIC,
from ideal hydrodynamics (solid curve) using the codes in 
\cite{Huovinen:2001cy,Kolb:2000sd} 
and covariant transport (squares) 
using the MPC algorithm \cite{MPC}. 
An initial (thermalization) time of $\tau_0 = 0.1$~fm$/c$ was assumed.
Transport results for a constant cross section (open squares) and for
$\eta/s \approx 1/(4\pi)$ (filled squares) are shown, 
while the hydrodynamic curve is from
\cite{hytrv2}.}
\label{fig:2}
\end{figure}

Figure~\ref{fig:2} shows differential elliptic flow $v_2(p_T)$ results
for $\tau_0 = 0.1$~fm. Even for a ``minimal''
$\eta/s = 1/(4\pi)$ (filled squares), dissipation reduces elliptic
flow at moderate $p_T \sim 2-3$ GeV by about $25$\% relative to the ideal
hydrodynamic limit (solid line). The relative 
change increases with decreasing $p_T$, and therefore dissipation 
also flattens 
the slope at low $p_T$%
\footnote{Even for the massless equation of state considered here,
at {\em very} low $p_T \lton 0.1$~GeV, $v_2(p_T) \propto p_T^2$ as 
follows from 
continuity and differentiability of the phasespace density. 
But the quadratic behavior very near the origin is 
masked by the linear rise in a 
wide  $p_T \sim 0.2 - 1.2$ GeV window.}.
For comparison, we also show the result (open squares) 
for a constant cross section $\sigma_{tr} = \sigma_0$, i.e., growing 
$\eta/s \approx (\tau/\tau_0)^{2/3} / (4\pi)$,
which of course generates much smaller elliptic flow.

\begin{figure}[htpb]
\epsfysize=6cm
\epsfbox{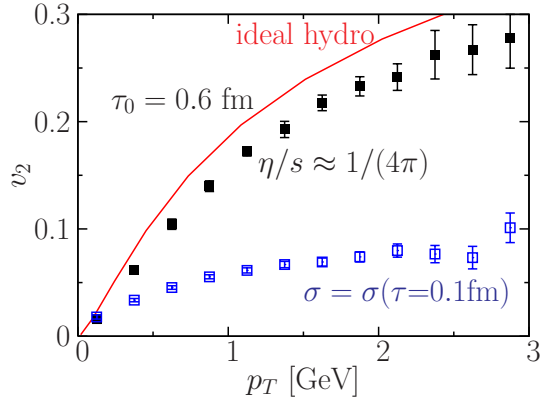}
\caption{Same as Fig.~\ref{fig:2} but for an initial (thermalization) time
$\tau_0 = 0.6$~fm$/c$.}
\label{fig:3}
\end{figure}

Figure~\ref{fig:3} compares differential elliptic flow $v_2(p_T)$
from the transport (squares) to the ideal hydrodynamic limit (solid line)
for a later thermalization time $\tau_0 = 0.6$~fm. In this case the initial
temperature was adjusted to $T_0 \approx 0.385$~GeV to account for cooling 
$T\sim \tau^{-1/3}$ because of $p\,dV$ work during longitudinal Bjorken 
expansion. With the rescaled temperature, results for both ideal hydrodynamics
and transport with {\em constant} cross section (open squares) 
are essentially independent of $\tau_0$, 
as long as $\tau_0 / R \ll 1$, as was also found in Ref.~\cite{hytrv2} 
(cf. Fig.~\ref{fig:2}).
However, with $\eta/s \approx 1/(4\pi)$, the 
dissipative reduction of elliptic flow (solid squares) relative to ideal
hydrodynamics is more modest,
$\sim 15$\%, for the larger $\tau_0 = 0.6$~fm. 
This is similar in magnitude to recent results from 
2+1D dissipative hydrodynamics\cite{romatschkeviscous,teaneyviscous}.

At fixed $\eta/s$, dissipative effects are weaker for larger $\tau_0$ 
because the initial value of the transport cross section 
is larger in that case.
An equivalent explanation is that though $\eta/s$ is the same, initial
velocity gradients in the system 
$\partial_\mu u_\nu \sim 1/\tau$  are smaller for larger $\tau_0$. Indeed, 
the Navier-Stokes correction 
to the stress tensor\cite{GyulassyDanielewicz85,DeGroot}
\be
T^{\mu\nu} = T^{\mu\nu}_{ideal} + 
\eta (\nabla^\mu u^\nu + \nabla^\nu u^\mu 
        - \frac{2}{3}\Delta ^{\mu\nu} \partial^\alpha u_\alpha)
\ee
($\Delta^{\mu\nu} \equiv g^{\mu\nu} - u^\mu u^\nu$, 
$\nabla^\mu \equiv \Delta^{\mu\nu} \partial_\nu$, $u^\mu$ is the
flow velocity), in case of a longitudinal boost-invariant expansion,
implies viscous corrections to the
transverse and longitudinal pressure
$\Delta p_T = 2\eta/(3\tau)$,
$\Delta p_L = -4\eta/(3\tau)$. 
Therefore, {\em relative} pressure corrections 
\be
\frac{\Delta p}{p} \sim \frac{\eta}{s} \frac{Ts}{p T\tau} 
= \frac{4\pi\eta}{s} \frac{1}{\pi T\tau}
\label{reldpNS}
\ee
decrease with time
$\Delta p / p \propto \tau^{-2/3}$, if
$\eta/s = const$; whereas $\Delta p / p \propto \tau^{0}$, 
if $\sigma_{tr} = const$.

Based on the Navier-Stokes estimate (\ref{reldpNS}),
it is not a surprise that dissipative corrections are important
for conditions expected at RHIC.
$\Delta p /p \sim 20$\% for $\tau_0 T_0 = 0.6$~fm $\times\ 0.385$~GeV
and is almost $100$\% for 
$\tau_0 T_0 = 0.1$~fm $\times\ 0.7$~GeV. In the former case,
the correction is modest, and viscous hydrodynamics is likely applicable.
In the latter case, however, a
hydrodynamic approach seems questionable.
It would be interesting to test this anticipated break-down of 
hydrodynamics against solutions of Israel-Stewart theory.
We note that dissipation is expected to be relevant not only at RHIC but 
at the LHC as well\cite{LHCv2,XGetas}.

{\em Conclusions.} We utilized covariant transport theory to 
study the effect of a ``minimal'' shear 
viscosity $\eta = s/(4\pi)$ on differential elliptic flow $v_2(p_T)$ in 
$Au+Au$ collisions at $\sqrt{s_{NN}} \sim 200$~GeV at RHIC. 
The key ingredient is a transport cross section 
$\sigma_{tr} \propto \tau^{2/3}$ that grows with time. 
We find significant reduction of elliptic flow relative to the ideal 
hydrodynamic limit, $\sim 25$\% reduction for a thermalization time 
$\tau_0 = 0.1$~fm, while $\sim 15$\% for $\tau_0 = 0.6$~fm.
This indicates that even such a small shear viscosity cannot be ignored
at RHIC, and thus the LHC as well, because gradients are large.

We note that this study set $\eta/s \approx 1/(4\pi)$ only within $20$\%
and in an average sense, for numerical convenience. The evolution of the 
local density
and temperature was approximated with analytic results for
longitudinal Bjorken expansion. More accurate results could 
be obtained, in principle, via a transport cross section $\sigma_{tr}(n,\tau)$
that depends explicitly on the local density. However, that
is much more expensive numerically with the solution technique
(cascade algorithm) utilized here.

In principle, radiative processes such as $gg\leftrightarrow ggg$ can also 
be included\cite{XGetas}.
However, near the hydrodynamic limit 
we do not expect large corrections to our results 
because the dynamics is determined solely 
by the equation of state and the viscosity $\eta/s = 1/(4\pi)$.
The main difference is that radiative processes allow for change
of particle number, which should be only a modest refinement for the
initial conditions considered in this study.

Finally, we 
emphasize that a simple ideal gas equation of state $e=3p$ has been 
considered, which also implies vanishing bulk viscosity.
It would be important to repeat this study with 
an equation of state that is 
more realistic for quark gluon matter at moderate
$T \lton 300$ MeV, and to investigate effects of bulk viscosity
which is expected to rise sharply in the vicinity of $T\sim 200$ 
MeV\cite{TKbulk}.

{\em Acknowledgments.}
I thank RIKEN, 
Brookhaven National Laboratory and
the US Department of Energy [DE-AC02-98CH10886] for providing facilities
essential for the completion of this work; 
and the hospitality of INT/Seattle, where a part of this work has been 
completed.

\end{document}